\documentclass[11pt]{article}
\pdfoutput=1
\usepackage{amsthm,graphicx}
\usepackage[letterpaper,margin=1in]{geometry}
\usepackage{moreverb,url,xcolor}
\usepackage[frozencache,cachedir=.]{minted}
\usepackage{blindtext}
\usepackage{marginnote}
\newcommand{\paragraphsummary}[1]{\marginnote{}}
\newcommand{\revision}[1]{#1}

\usepackage{hyperref}
\hypersetup{breaklinks=true, colorlinks=true, linkcolor={blue!90!black}, citecolor={blue!90!black}, urlcolor={blue!90!black}}
\usepackage{doi,authblk,cite}

\usepackage[nameinlink]{cleveref}

\def\volumeyear{2024}
\newcommand{\keywords}[1]{Keywords: \textit{#1}}

\begin{document}

\title{Portable, heterogeneous ensemble workflows at scale using libEnsemble}

\author[1]{Stephen Hudson}
\author[1]{Jeffrey Larson}
\author[1]{John-Luke Navarro}
\author[2,3]{Stefan M.~Wild}
\affil[1]{MCS, Argonne National Laboratory}
\affil[2]{AMCR, Lawrence Berkeley National Laboratory}
\affil[3]{IEMS, Oak Ridge National Laboratory}
\affil[ ]{~\linebreak\texttt{shudson@anl.gov}}

\maketitle

\begin{abstract}
libEnsemble is a Python-based toolkit for running dynamic ensembles, developed
as part of the
DOE Exascale Computing Project.
The toolkit utilizes a unique generator--simulator--allocator paradigm, where generators
produce input for simulators, simulators evaluate those inputs, and allocators
decide whether and when a simulator or generator should be called. The
generator steers the ensemble based on simulation results. Generators may,
for example, apply methods for numerical optimization, machine learning, or
statistical calibration.

libEnsemble communicates between a manager and workers. Flexibility
is provided through multiple manager--worker communication
substrates each of which has different benefits. These include Python's
multiprocessing, mpi4py, and TCP. Multisite
ensembles are supported using Balsam or Globus Compute.

We overview
the unique characteristics
of libEnsemble as well as current and potential interoperability with other
packages in the workflow ecosystem.
We highlight libEnsemble's dynamic resource features:
libEnsemble can detect system resources, such as available nodes, cores, and GPUs,
and assign these in a portable way. These features allow users to specify the number of processors and GPUs required for
each simulation; and resources will be automatically assigned on
a wide range of systems, including
Frontier, Aurora, and Perlmutter.
Such ensembles can include multiple simulation types, some using GPUs
and others using only CPUs, sharing nodes for maximum efficiency.
We also describe the benefits of libEnsemble's generator--simulator coupling,
which easily exposes to the user
the ability to cancel, and portably kill, running simulations based on models that are updated with intermediate simulation output.

We demonstrate libEnsemble's capabilities, scalability, and scientific impact
via a Gaussian process surrogate training problem for the longitudinal density profile at the exit of a plasma accelerator stage.
The study uses gpCAM for the surrogate model and employs either Wake-T or WarpX simulations, highlighting efficient use of resources
that can easily extend to exascale.
\end{abstract}


\section{Introduction}

Extreme-scale scientific computing experiments and applications enjoy much success from increasing resource and scaling capabilities. \revision{ However, the vast majority of practical simulations have a limit to which they can scale at a reasonable parallel efficiency. Running ensembles presents an effective way to utilize these increasing resources towards a unified goal.}

libEnsemble\cite{Hudson2022,Hudson2023,libE,libEnsembleManual} is a Python workflow toolkit that coordinates \textit{ensembles}:
concurrent instances of calculations or applications at the quantities and scales possible on modern supercomputers. But
running \textit{naive} ensembles of computations from predetermined input parameters (e.g., via a fancy
\texttt{for} loop) is often not the most efficient route toward a study's goals. Therefore, libEnsemble
coordinates \textit{dynamic ensembles}, where ensemble members are produced and
controlled on the fly, without human interaction, based on the instructions of some outer-loop, model, or decision processes.

libEnsemble is aware of the heterogeneous resources (CPUs and GPUs) available on 
many target machines (including laptops, clusters, and supercomputers), and automatically detects, allocates,
and reallocates such compute resources.
For a given ensemble, the decision, allocation, and simulation components are selected or supplied to
libEnsemble in the form of \textit{user functions}, described below.

libEnsemble is one of several workflow packages addressing the need to reliably and portably scale
concurrent computations across a landscape of heterogeneous hardware. For example, the RADICAL-Cybertools Ensemble
Toolkit\cite{ensembletoolkit} enables the coordination of ensembles of simulations that adhere to common patterns.
Many workflow packages, including Colmena\cite{colmena} and Ray,\cite{ray} specifically target artificial intelligence use cases. Covalent\cite{covalent} encourages
extremely portable, cross-site workflows targeting both clusters and cloud-computing providers. libEnsemble is 
distinguished from each of these other examples (except Colmena) by only requiring a user to select/supply \textit{at most}
three components (generators, simulators, and allocators), not requiring a directed acyclic graph of tasks, not requiring administrating background processes or databases, and supporting
dynamic assignment/reassignment of compute resources. Colmena offers a similar two-component paradigm (using the terms ``thinkers'' and ``doers'')
but requires more infrastructure configuration;
libEnsemble's users typically do not need to specify data pipelines, queues, databases, or other interoperative glue.

libEnsemble was born out of discussions with the PETSc/TAO\cite{petsc-user-ref320} team centering on \revision{the need for modular workflow libraries to support dynamic ensembles to help} the U.S.~Department of Energy (DOE) to realize the potential of exascale computing capabilities. \revision{It was determined that the library should have a mix and match approach to generator and simulator functions, should allow for a flexible and high resolution resource partitioning (nodes and sub-nodes), and would allow efficient use of resources by being able to direct the termination of running simulations. While the TAO optimizers were a focus of discussion, the library should support any kind of mathematical function to guide simulations.} Initiating development under the DOE Exascale Computing Project, today
libEnsemble plays an active role within the DOE and broader workflows community, featuring integrations with prominent ExaWorks packages, including PSI/J\cite{10254912} and Globus Compute.\cite{funcx} Its presence extends to the mathematical software community as a member of the Extreme-scale Scientific Software Development Kit (xSDK\cite{bartlett2017xsdk}). Additionally, libEnsemble contributes to the quality assurance of scientific software by participating in the test suite of the Extreme-scale Scientific Software Stack (E4S).
These capabilities place libEnsemble at the forefront of facilitating scalable, resource-aware scientific computing, enabling intricate simulations and analyses across various scientific domains.

\subsection{Illustrative Use Cases}

\subsubsection{Research:}

Many uses of libEnsemble center on its ability to empower domain
scientists to utilize additional computation when their simulation
stops scaling. As discussed, if a simulation's run time does not decrease with
additional resources, libEnsemble exposes an additional opportunity for
parallelism by concurrently evaluating multiple simulations under different input parameters.
Countless examples of such computational situations exist. A particular set of exemplars is how libEnsemble
has been beneficial to the particle accelerator simulation
community. For example, Neveu et al.\cite{Neveu2019,Neveu2022} applied libEnsemble to optimize photoinjectors,
while Ferran Pousa et al.\cite{FerranPousa2022} performed multifidelity optimization of 
laser-plasma accelerators.

\subsubsection{Generators:}

While researchers often create or customize libEnsemble generators for particular workflows, some generators or suites of generators may be developed and shared by multiple researchers.
Examples of community-developed generators include the VTMOP\cite{Chang2020} generator for multiobjective optimization,
the Surmise\cite{surmise2023,chan2023high} generator for uncertainty quantification via Gaussian processes, and the ytopt\cite{Wu2024}
generator for machine-learning-based autotuning.

Further examples can be found in the libEnsemble community examples repository on GitHub.\cite{libEnsembleCommunityExamples}

\subsubsection{Software:}

Optimas\cite{PhysRevAccelBeams.26.084601} is a package for highly scalable parallel optimization by Ferran Pousa et al.
It is built on top of libEnsemble and offers a higher-level API that is designed for accelerator modeling but applicable more broadly. Optimas development is closely aligned with libEnsemble's development.

ParMOO\cite{parmoo,ParMOODesign23} provides a library of multiobjective optimization solvers that is designed to integrate with libEnsemble for running evaluations across parallel resources. The libEnsemble developers continue to work with optimization library teams to provide a standardized, modular environment for dynamic workflows on a wide range of parallel platforms.

rsopt~\cite{rsopt} is another Python framework built around libEnsemble. It is used for testing and running black-box optimization problems. Developed by RadiaSoft, rsopt is in turn a key element in Sirepo, a proprietary interface for accessing
and running jobs on high-performance computing systems via a web browser interface.

\section{Manager, Workers, and User Functions}

\begin{figure}
    \centering
    \includegraphics[width=2in]{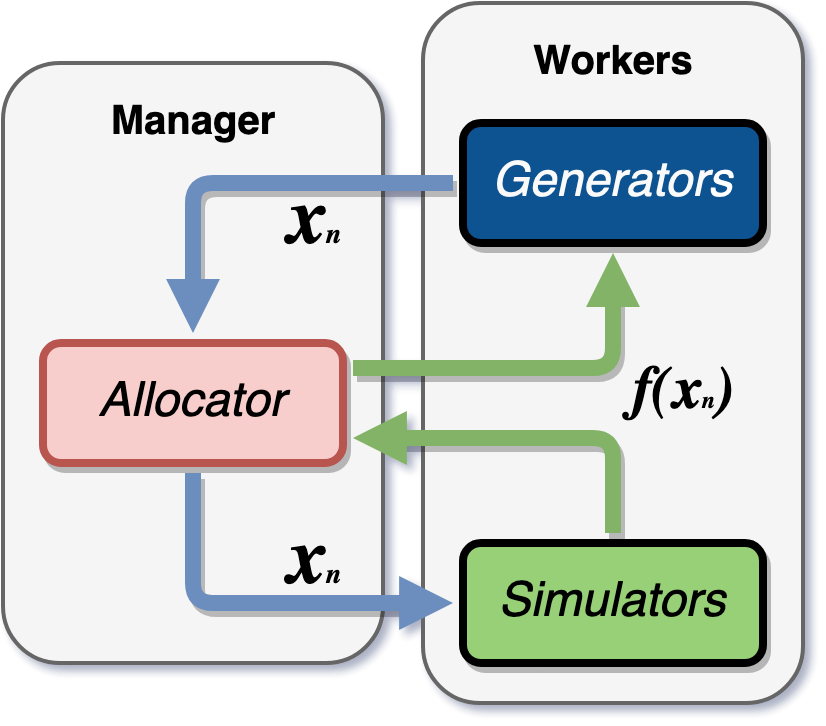}
    \caption{Interoperability among libEnsemble's allocator, generators, and simulators user functions}
    \label{fig:Interoperability}
\end{figure}

libEnsemble's workflows are driven by the interoperability between three components
referred to as \textit{user functions}, coordinated by a \textit{manager}
and multiple \textit{workers}. \textit{Generator}
user functions produce candidate inputs/parameters for simulations or experiments.
\textit{Simulator} functions perform and monitor those simulations or experiments.
\textit{Allocator} functions coordinate data transfer between the
generators and simulators and can additionally assign compute resources to user
functions or perform additional tasks such as cancelling active simulations based on a request
from the generator. libEnsemble's workers typically launch
generator and simulator functions, while the manager handles the allocator function, as depicted in \Cref{fig:Interoperability}.
Thousands of concurrent simulator and generator
instances can be asynchronously coordinated and tightly coupled together in a namesake
\textit{ensemble}.

User functions are simply Python functions that conform to libEnsemble's API; that is,
they must process their first input parameter as a NumPy\cite{harris2020array}  structured
array and then return/communicate another array containing results. Between these two steps, any level of
computation and complexity is possible. Additional Python libraries can be
imported, multinode compiled applications launched and monitored via libEnsemble's built-in executors,
models trained, or any other routine performed. With few exceptions, any kind of
computation is possible in libEnsemble's user functions.

User functions commonly have a state that they wish to 
maintain and update as results are observed. 
Such functions can be launched in libEnsemble in a persistent mode, as opposed to a fire-and-forget mode 
that is commonly used in other workflow packages.
This persistent mode is facilitated by user-facing 
\textit{persistent support} routines that allow user functions to continuously send
and receive intermediate results or additional input parameters while maintaining
state.
For example, APOSMM,\cite{Larson2018} a parallel optimization generator function distributed with libEnsemble,
persistently maintains and advances multiple local-optimization subprocesses as simulation
results are returned to APOSMM in real time.

libEnsemble distributes dozens of example functions and complete workflows as part of its package and online community examples.\cite{libEnsembleCommunityExamples}
Users are welcome to browse these examples, distribute or modify them to serve their purposes, and even contribute new examples.

\subsection{History Array}

libEnsemble's manager maintains a NumPy structured array referred to as the
\textit{history array}, which contains a record of all generated and evaluated
values and corresponding metadata. Results from each user function are slotted
into this array, and inputs into user functions are selected from it. Upon
completing a workflow, incrementally, or encountering an error, libEnsemble
dumps the history array to file.

Generator functions can preempt, prioritize, or otherwise
manipulate the run order of simulations by assigning \texttt{sim\_id}s 
and \texttt{priority} rankings to corresponding output parameters in the history array.
\revision{Restarts are supported by supplying a history array from a previous ensemble in the \texttt{H0} argument when the ensemble is executed.}

\subsection{Generator Functions}

Generator functions in libEnsemble play a pivotal role in defining the
exploration strategy within the parameter space for simulations or experiments.

Since the possible objectives driving explorations are countless, corresponding generators
range from simple designs, such as those requesting batches of randomly sampled points, to
substantially more complex methods. For instance, APOSMM coordinates multiple instances of
structure-exploiting numerical optimizers. Each optimizer instance may begin its search
from a distinct point in the parameter space, leveraging the parallel computing capabilities of
libEnsemble to  efficiently explore large and complex parameter spaces for simulations that are
likely already using (considerable) parallel resources. 

The flexibility of generators allows users to tailor their search
strategy according to the problem at hand, whether it involves scanning a broad
area to identify regions of interest or focusing on refining solutions within one.
Moreover, generator functions can 
dynamically adjust their requests based on real-time output 
from ongoing simulations. This includes the capability to request
the cancellation of currently running
simulations if, for example, those simulations are deemed unlikely to give useful output. 
Cancellation could also be useful if robustness to perturbations in input parameters is desired 
but there are running simulations using input parameters that are very close to those that have
been discovered to crash the simulation.
Such cancellation functionality is
critical when managing expensive-to-evaluate simulations and computational resources are limited. libEnsemble has been used in this way to limit the number of expensive energy density functional evaluations performed, which has led to additional generator methodology research on how to address such missing output; see, for example, Chan et al.\cite{MCMPSW2022}

\begin{figure}[h!]
\centering
\begin{minted}[
frame=lines,
fontsize=\scriptsize,
]
{python}
import numpy as np
from simple_models import Model
from libensemble.message_numbers import *

from libensemble.tools.persistent_support import (
    PersistentSupport,
)


def persis_gen(H_in, persis_info, gen_specs, libE_info):
    us = gen_specs["user"]
    b, lb, ub = us["gen_batch_size"], us["lb"], us["ub"]
    n = len(lb)

    x = persis_info["rand_stream"].uniform(lb, ub, (b, n))
    H_o = np.zeros(b, dtype=gen_specs["out"])
    H_o["x"] = x

    model = Model(H_in["x"], H_in["f"], bounds=(lb, ub))
    ps = PersistentSupport(libE_info, EVAL_GEN_TAG)
    tag, Work, calc_in = ps.send_recv(H_o)

    while tag not in [STOP_TAG, PERSIS_STOP]:
        model.update(x, calc_in["f"])
        npoints = len(calc_in)
        x = model.ask(npoints)

        H_o = np.zeros(npoints, dtype=gen_specs["out"])
        H_o["x"] = x
        tag, Work, calc_in = ps.send_recv(H_o)

    return None, persis_info, FINISHED_PERSISTENT_GEN_TAG

\end{minted}
\caption{Persistent generator function using a model to decide new simulation inputs. \label{fig:simple_gen}}
\end{figure}

\Cref{fig:simple_gen} illustrates a simple persistent generator that updates a model with simulation outputs and decides new inputs for future simulations.

The generator first unpacks fixed parameters given by the manager. It then starts by producing \texttt{b} points in
the $n$-dimensional box defined by the vectors \texttt{lb} and \texttt{ub} using the given seeded random stream. The model, which could be any external package, is initialized with \texttt{H\_in} if previous results are being used (for example this may be a restart), otherwise this is an empty array. The main loop updates the model with simulation results and generates new points for evaluation. The \texttt{PersistentSupport} class provides communication functions to send new points to the manager and receive results of simulations, where the \texttt{ps.send\_recv} function blocks on the receive.

\subsection{Simulator Functions}

\begin{figure}[!t]
\centering
\begin{minted}[
frame=lines,
fontsize=\scriptsize,
]
{python}
import numpy as np

def sim_f(H, persis_info, sim_specs):
    x = H['x']
    H_o = np.zeros(1, dtype=sim_specs['out'])
    H_o['f'] = np.linalg.norm(x)
    return H_o
\end{minted}
\caption{Simple simulator function\label{fig:simple_sim}}
\end{figure}

The outputs from the generator functions are usually the inputs for the
simulator functions. The generator-produced inputs could be meshes used for a
computational fluid dynamics simulation, magnet strengths and locations for a
synchrotron particle accelerator,
or hyperparameters for a convolutional neural network. Simulations can use purely
CPU or  CPU and GPU resources to perform their
computations. Simulator functions can be extremely simple, such as the one in 
\Cref{fig:simple_sim}, or arbitrarily complex and depend on
external executables or non-Python libraries. To make it as easy as possible to
interface with such simulators, the libEnsemble executor 
can be used to launch applications and monitor results. 

\subsection{Calling libEnsemble}

\revision{

libEnsemble's configuration can use the original dictionary and function interface or the more recent object based interface. An example configuration and execution of an ensemble using the object interface is illustrated in \Cref{fig:calling_script} (imports are excluded for brevity). This script specifies the configuration for the generator, simulator, and allocator functions, defines the exit condition (after 500 simulations), provides random streams to each worker, and runs the ensemble.  The inputs supplied are automatically validated using Pydantic \cite{pydantic}.}

\begin{figure}[t]
\centering
\begin{minted}[
frame=lines,
fontsize=\scriptsize,
]
{python}
libE_specs = LibeSpecs(nworkers=4)

sim_specs = SimSpecs(
    sim_f=sim_f,
    inputs=["x"],
    outputs=[("f", float)],
)

gen_specs = GenSpecs(
    gen_f=persis_gen,
    inputs=["x", "f"],
    persis_in=["f"],
    outputs=[("x", float, 2)],
    user={
        "gen_batch_size": 50,
        "lb": np.array([-3, -2]),
        "ub": np.array([3, 2]),
    },
)

alloc_specs = AllocSpecs(alloc_f=only_persistent_gens)
exit_criteria = ExitCriteria(sim_max=500)

ensemble = Ensemble(
    libE_specs=libE_specs,
    sim_specs=sim_specs,
    gen_specs=gen_specs,
    alloc_specs=alloc_specs,
    exit_criteria=exit_criteria,
)

ensemble.add_random_streams()
H, persis_info, flag = ensemble.run()
\end{minted}
\caption{Configuring and running an ensemble.\label{fig:calling_script}}
\end{figure}

\subsection{Application Launchers -- Executors}

libEnsemble features \textit{executors} for portably launching and monitoring
user applications. libEnsemble's
\textit{MPIExecutor} builds
MPI run lines based on compute-resource counts, MPI distribution information, scheduler
information, and user inputs. This executor is coupled to libEnsemble's comprehensive
resource detection and allocation. For example, an \texttt{Executor.submit()}
instruction to launch an application across thirty-two GPUs will not
need any adjustment whether libEnsemble is running on Intel, AMD, or Nvidia resources, with
any common scheduler or MPI distribution, and irrespective of the number of GPUs per node. \revision{An example is given in \Cref{fig:executor_in_sim} where all GPUs assigned to the worker will be used. While \Cref{fig:gen_set_gpus} shows code from a generator functions that sets the number of GPUs based on the size of the problem to be evaluated.}

\begin{figure}[h!]
\centering
\begin{minted}[
frame=lines,
fontsize=\scriptsize,
]
{python}
from libensemble.executors import MPIExecutor

# Initialize MPI Executor
exctr = MPIExecutor()
sim_app = os.path.join(os.getcwd(), "../forces.x")
exctr.register_app(full_path=sim_app, app_name="forces")

ens = Ensemble(executor=exctr...
\end{minted}
\begin{minted}[
frame=lines,
fontsize=\scriptsize,
]
{python}
def run_forces(H, persis_info, sim_specs, libE_info):
    """Launches the forces MPI app.
    
    Assigns one MPI rank to each GPU assigned to worker.
    """

    # Parse out num particles and set args
    particles = str(int(H["x"][0][0]))
    args = particles + " " + str(10)

    # Retrieve our MPI Executor
    exctr = libE_info["executor"]

    # Submit our forces app for execution.
    task = exctr.submit(
        app_name="forces",
        app_args=args,
        auto_assign_gpus=True,
        match_procs_to_gpus=True,
    )

    task.wait()

    data = np.loadtxt("forces.stat")
    final_energy = data[-1]

\end{minted}
\caption{\revision{Use of MPI Executor in calling script (above) and simulation function (below). These scripts will work on various systems including Perlmutter, Frontier, and Aurora.}\label{fig:executor_in_sim}}
\end{figure}

\begin{figure}[t]
\centering
\begin{minted}[
frame=lines,
fontsize=\scriptsize,
]
{python}

while tag not in [STOP_TAG, PERSIS_STOP]:
    x = rng.uniform(lb, ub, (b, n))
    bucket_size = (ub[0] - lb[0]) / max_gpus

    ngpus = [
        int((num - lb[0]) / bucket_size) + 1
        for num in x[:, 0]
    ]
    H_o = np.zeros(b, dtype=gen_specs["out"])
    H_o["x"] = x
    H_o["num_gpus"] = ngpus

\end{minted}
\caption{\revision{Generator code that sets the number of GPUs for a given evaluation based on particle count, at the same time as setting evaluation inputs.} \label{fig:gen_set_gpus}}
\end{figure}

libEnsemble's executors also feature \texttt{poll}, \texttt{kill},
\texttt{wait}, and other functions for monitoring the status of
applications. The \texttt{manager\_poll} function, in particular, checks for signals
relayed by the manager from a generator, enabling generator functions to
request the cancellation and termination of already-running applications.

\revision{The MPIExecutor interacts with the detected system scheduler (SLURM, PBS, Cobalt,
LSF) to distribute application runs over available nodes. The built-in scheduler
will attempt to minimise the number of nodes used for a given resource requirement
when only \texttt{num\_procs} and/or \texttt{num\_gpus} are specified in the generator,
but will find available slots across multiple nodes as necessary.

The MPIExecutor \texttt{submit()} function allows the user to directly specify portable
options such as \texttt{num\_procs}, \texttt{num\_nodes}, \texttt{num\_gpus}, and
\texttt{procs\_per\_node}, while also allowing any customized string of MPI runner
options to be added as \texttt{extra\_args}. In some cases, user scripts read \texttt{extra\_args} from
an environment variable in the batch script to maintain portability of scripts. An \texttt{env\_script} argument will instead construct a bash script that can set enviornment variables, load modules, and runs in the sub-process without affecting the worker's own environment.}

\begin{figure}[h!]
    \centering
    \includegraphics[width=3in]{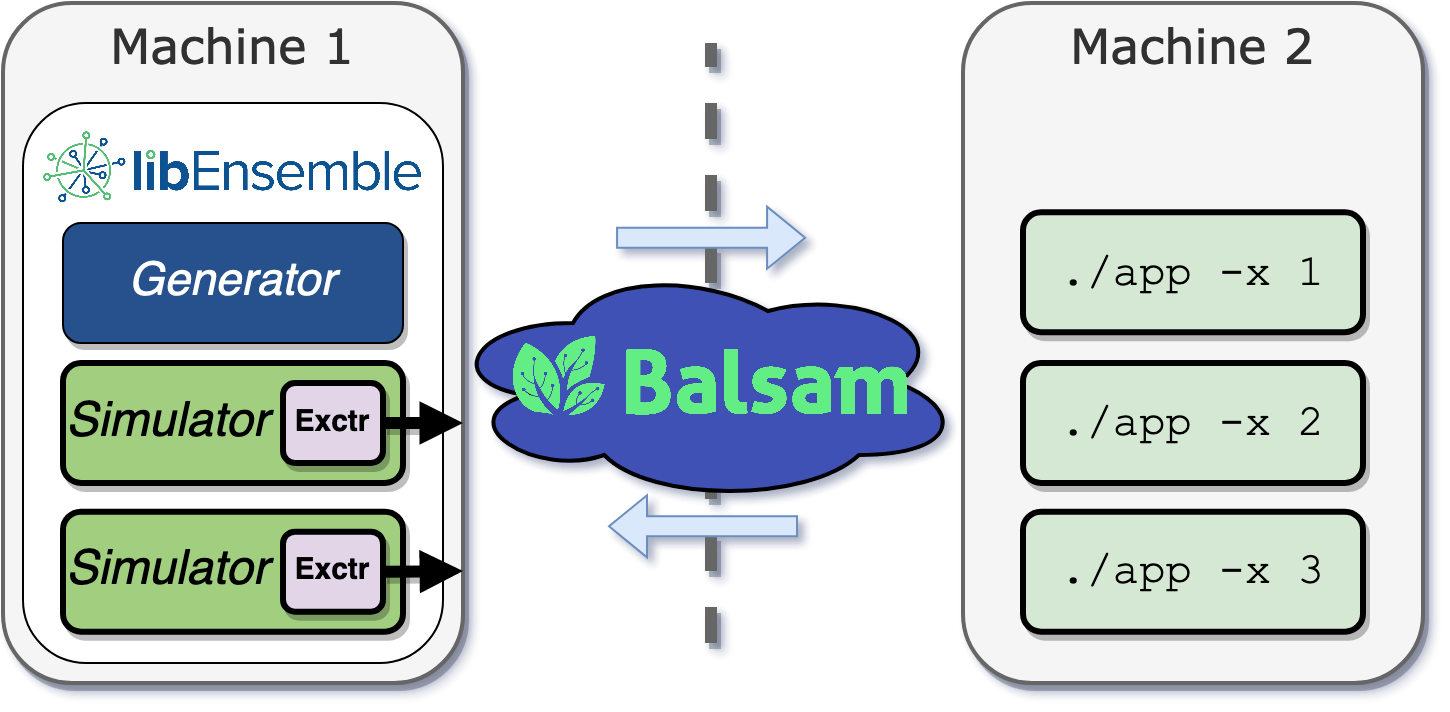}
    \caption{Using Balsam in libEnsemble to run applications across machines}
    \label{fig:balsam}
\end{figure}
The \textit{MPIExecutor} can be swapped out with the
\textit{BalsamExecutor} so simulation functions can submit application
instances onto separate resources, as depicted in \Cref{fig:balsam}, enabling cross-site
heterogeneous resource ensembles. This process
additionally involves wrapping the application command in Balsam's \texttt{ApplicationDefinition}
class, then running a \texttt{BalsamExecutor.submit\_allocation} request to reserve
compute resources on the remote system. This system is not tightly coupled to libEnsemble's
local resource management, but the BalsamExecutor can still request variable amounts of
compute resources from remote systems, ensuring that dynamically sized application launches are
possible.

\revision{As user functions are written in Python and run in a worker-specific process, it is easy for a user to express a multi-component workflow. For example, one or more application runs may be used via the executor, and analysis performed using Python tools.}

\subsection{Allocator Functions}

The allocation function orchestrates the ensemble, deciding whether simulation
or generation work should be performed as the ensemble progresses, possibly
considering the outputs from intermediate calculations. The simplest (and
default) libEnsemble allocation function calls the generator function when no
simulation work exists and otherwise sequentially gives out previously
generated simulation work that has not started to be simulated. While
sophisticated generators can observe previous simulation results and adjust their
future requests, not all approaches are so advanced. In such cases, the allocation
function can have logic to perform such capabilities. The worker array and the
history array are central to the allocation functions' decisions. The worker
array gives a snapshot of the currently running activities and the available
resources. The history---containing all pending and completed computations
performed by workers, including their outputs and run times---enables the
allocation function to know the state of the ensemble and to dynamically make
decisions about future work. This dynamic decision-making process ensures that
the computational resources are used efficiently, prioritizing tasks that are
most likely to advance the ensemble's objectives. It allows for real-time
adaptation as the knowledge of the simulation's landscape improves. 

\subsection{Simulation Cancellation}

\revision{The tightly coupled nature of user functions within libEnsemble allows generators
to dynamically cancel simulations. This capability is especially valuable when computations are expensive since computational resources may then
be used on more critical evaluations as determined by a generator. For example, a recently updated surrogate model may predict that a previously issued evaluation will not be informative based on results that have returned in the meantime; in this case, the previously issued evaluation can be cancelled, with freed resources made available for other evaluations.

Cancelled simulations that have not yet been issued to a worker will not be issued. Also, if
the \texttt{kill\_canceled\_sims} feature is enabled, kill signals are sent to workers with currently
running simulations. These are caught by workers that can then terminate simulations and perform
any necessary cleanup.
libEnsemble includes helper routines such as \texttt{Exectutor.polling\_loop()} to simplify this process. When run in a simulator function, the routine polls for kill signals from the manager and uses a portable termination method to end the running simulation.}

Assigned computational resources are
then recovered and reallocated to more valuable computations.

\subsection{Running on Multiple Compute Sites}

libEnsemble features three approaches for running cross-machine ensembles of computations.

\subsubsection{Balsam:}

Via Balsam,\cite{balsam} libEnsemble's user function instances can 
schedule and launch \textit{applications} on separate machines, as depicted in \Cref{fig:balsam}.
This approach separates workflow and application compute resources. User functions can also
request and relinquish compute resources on the separate machine for those applications.

\subsubsection{Globus Compute:}

Via Globus Compute (formerly funcX\cite{funcx}), libEnsemble's worker processes can submit their \textit{user functions}
to separate machines. This allows fire-and-forget variants of user functions (usually
simulators) to themselves use parallel resources such as MPI or be co-located with their
launched applications on the separate machine.

\subsubsection{Reverse SSH:}

Via a lower-level reverse-SSH interface, libEnsemble can launch its \textit{worker processes}
onto separate machines, effectively splitting the workflow software itself across a network. Persistent
user functions launched by remote workers can communicate with libEnsemble's manager across
the network, unlike with Globus Compute; but fewer concurrent worker processes are formally supported.

\section{Resource Detection and Management}

\revision{When a job scheduler is present, libEnsemble typically runs in a single batch submission. libEnsemble queries the scheduler for available nodes, and by default,} equally divides available compute resources among workers. Hence, if users are running 
fixed-sized simulations, they do not need to consider dynamic resources. However, libEnsemble has built-in support 
for dynamic resource allocation. 
\revision{Here, dynamic refers to the ability of the user to assign different compute resources (including CPU cores and GPUs) to each evaluation.}

To support portability, libEnsemble incorporates considerable detection of
system information, including scheduler details, MPI runners \revision{(\textit{mpirun}, \textit{srun}, \textit{jsrun}, \textit{aprun})}, nodes, core counts, and
GPU counts, and uses these to produce run lines and GPU settings for these
systems, without the user having to modify scripts. For example, libEnsemble can
detect NVIDIA, AMD, and Intel GPUs and uses the detected system information, along with MPI
runner/scheduler information, to assign work in an appropriate way for the system, be
it command line settings or environment variables such as
\texttt{ROCR\_VISIBLE\_DEVICES} or \texttt{CUDA\_VISIBLE\_DEVICES}. This means
that when simulation input parameters are created, in the generator function, the number of processes and GPUs
for each simulation can also be set, and libEnsemble will assign resources correctly for
the system.

\revision{
Internally, libEnsemble uses a concept of resource sets, where each resource set consists of CPU resources divided by the number of workers. GPU resource sets are also separately maintained, ensuring that CPU-only applications do not prevent GPU usage by other workers. Note that each resource set may range from a sub-node partition to a block of multiple nodes. When users specify the number of processors and/or GPUs required for an evaluation, this is internally converted to resource sets, which are then assigned to each worker. This means users do not need to modify scripts to account for the number of processors or GPUs per node---this is all managed by the internal resource scheduler. 

These capabilities constitute an update since the publication of Ref.~\citenum{Hudson2022}. Previously, users had to assign the number of resource sets directly, and also had to explicitly set the appropriate GPU settings for the system they were using (e.g., \texttt{CUDA\_VISIBLE\_DEVICES} or SLURM's \textit{srun} command line options such as \texttt{--task-per-node}). This automation has significantly simplified the user experience.
}

In some circumstances, users may wish to override detected settings. For example, there may be ambiguities, such as the presence of multiple MPI runners, or user preferences, such as alternative ways of assigning devices, or whether to treat GPU tiles as separate devices. Consequently, users have the option to specify platform settings with a number of documented fields. The approach taken by libEnsemble is to  honor explicitly provided settings while automatically detecting any that are not specified. A list of known platforms, including DOE leadership-class machines, is maintained within
libEnsemble; and these can also be specified in user scripts or via an environment variable. However, most of these known systems are also detected automatically. \revision{As an example, the default settings for Aurora and Frontier at time of writing are shown below.}

\begin{minted}[
frame=lines,
fontsize=\scriptsize,
]
{python}
class Aurora(Platform):
    mpi_runner: str = "mpich"
    runner_name: str = "mpiexec"
    cores_per_node: int = 104
    logical_cores_per_node: int = 208
    gpus_per_node: int = 6
    tiles_per_gpu: int = 2
    gpu_setting_type: str = "env"
    gpu_setting_name: str = "ZE_AFFINITY_MASK"
    scheduler_match_slots: bool = True

class Frontier(Platform):
    mpi_runner: str = "srun"
    cores_per_node: int = 64
    logical_cores_per_node: int = 128
    gpus_per_node: int = 8
    gpu_setting_type: str = "runner_default"
    gpu_env_fallback: str = "ROCR_VISIBLE_DEVICES"
    scheduler_match_slots: bool = False
    
\end{minted}

\revision{This determines that Aurora will use the environment variable \texttt{ZE\_AFFINITY\_MASK} to set GPUs. Frontier will use the \texttt{srun} defaults when using an MPI runner, but use \texttt{ROCR\_VISIBLE\_DEVICES} otherwise. The user can override any of these settings.
}

\subsection{Resource Scheduling}

\revision{The resource management component of libEnsemble includes a scheduler class that is used to examine available resources sets (with corresponding nodes/CPU/GPUs) and assign these to workers to meet user requirements. The scheduler attempts to fit simulations onto a node, but uses an even split across nodes if necessary. It can also support options \texttt{split2fit} which allows resource sets to be split across more nodes if space is not currently available on the minimum node count required (this is the default), and \texttt{match\_slots}, which ensures matching slot IDs (partitions within nodes), which is necessary when setting GPUs via an environment variable, which also defaults to True. Users also have the option of providing an alternative scheduler, and in the future, if more refined resource management is needed, a resource scheduler based on Flux \cite{flux-sched} may be incorporated for greater benefits.}

\subsection{Multiapplication Ensembles}

The dynamic resource approach allows ensembles that have multiple applications,
each using different resources. The multifidelity studies\cite{PhysRevAccelBeams.26.084601,FerranPousa2022} use many fast Wake-T simulations, running on
a single CPU core, alongside fewer, highly targeted particle-in-cell (PIC) simulations (e.g., using
FBPIC or WarpX), which may use one or more GPUs. This approach has demonstrated
a factor of 10 speedup in parameter optimization over the single-fidelity
approach. Nodes combine runs of Wake-T and FBPIC, allowing highly efficient
resource utilization on each node. While GPU applications often leave many CPU
cores idle, the additional CPU cores in this case are used to run Wake-T
simulations. Thus a double win is achieved: efficiency is gained by minimizing
expensive simulations, and resource efficiency is gained by exploiting the excess
CPU cores.

\section{Parallelization Methods}

libEnsemble supports four methods for initializing parallelization and manager/worker communications.
Each method offers various scaling/speed/local-data benefits and trade-offs, and libEnsemble workflows
are portable across each method.

\subsection{MPI}

Running with MPI-based processes
typically scales the best, is the most flexible with process placement on multinode
systems, and is often the most familiar to scientific users. Launching libEnsemble with MPI is
as simple as \\
\indent\texttt{\footnotesize mpirun -n 65 python my\_libensemble\_workflow.py},\\
which will start one manager
process on rank 0 with the remaining being reserved for workers. Other related MPI runners on clusters, such as \texttt{aprun} or \texttt{srun}, can also be used. Internally libEnsemble coerces runtime
information and communicates via \texttt{mpi4py}\cite{mpi4py} methods.

\subsection{Multiprocessing}

Referred to as \textit{local} communications, the multiprocessing mode uses Python's standard library \texttt{multiprocessing} module to
start separate Python interpreters/processes for workers, then communicate via
\textit{queues} from the same module. Each process is initialized on the same node, either a launch node
or a single compute node. The remaining nodes allocated within a job are accessible via libEnsemble's
executors for launching applications (workers and applications are \textit{not} co-located).

\subsection{Threading}

The threading mode starts separate threads for each worker from Python's standard library \texttt{threading} module. This
communication mode has the same locality limitations as \textit{multiprocessing}. Other major limitations
are that the resource allocation process is not thread safe and workers cannot maintain separate
working directories.

The benefits are that workers can both access shared data structures and communicate via reference; this mode
is significantly faster and ensures that models or other objects supplied to workers are updated
in place (not copied). For example, a generator's state is updated in place and is immediately accessible
to the manager.

We note that Python-level code is not technically parallel because of Python's
global interpreter lock (GIL). However, libraries such as NumPy, used throughout libEnsemble, offload computational heavy lifting to compiled code that bypasses the GIL.

Because of the limitations, this communications mode is not generally recommended, but it has proven useful to some users and has been used in live experiments, where libEnsemble is run through a Jupyter notebook.

\subsection{Manager-Run Generator}

A recent addition to libEnsemble is the \texttt{gen\_on\_manager} option that runs a generator as a thread on the manager. Previously, generator functions were always the responsibility of the worker processes to launch and handle. However, for common use cases involving only a single generator instance, placing that generator on a
worker process introduces inherent communication and process overheads that could be sidestepped by keeping that generator local to the manager process. While both inline and threaded methods have been evaluated, using a thread enables users to keep existing persistent generator functions and allows the main loop to keep running.

Furthermore, user input data is available for the generator to read/write by reference, exposing
partial benefits of libEnsemble's \textit{threading} parallelism to the \textit{MPI} and \textit{multiprocessing} methods. Since only one thread is run, the thread-safety issues between workers is not encountered, and many users find it more intuitive to set the number of workers to the number of parallel simulations.

\subsection{TCP}

TCP mode starts separate interpreters on SSH-accessible target systems for each worker
via a reverse-SSH interface. Workers communicate back to libEnsemble's manager
via a \texttt{multiprocessing.Manager}. This mode is a lower-level cross-system
ensemble approach compared with other solutions such as Globus
Compute; similarly, adaptive resource management is not applicable. However, unlike other current cross-site solutions, persistent user functions can be used remotely.

\section{Case Study: Online Learning of a Plasma Accelerator}

We now use libEnsemble to construct a surrogate model of a laser-plasma interaction using domain-specific 
simulation codes (WarpX and Wake-T), specifically focusing on the shape of a density down-ramp profile of a
plasma accelerator stage over a five-dimensional parameter space. This approach
capitalizes on the scalable capabilities of libEnsemble due to the arbitrary
concurrency available when exploring a parameter space.
The plasma accelerator use case being modeled features a Gaussian density bump as a plasma lens to focus the 
beam.\cite{Thaury2015} The objective function is a measure of beam divergence.

For the generator we use gpCAM,\cite{noack_2023_10393189} which 
implements function approximation and optimization using a Gaussian process\revision{, which is} a type of statistical surrogate model.\cite{gramacy2020surrogates} 
gpCAM is a highly customizable package for autonomous experimentation\cite{NoackUshizimaBook2023} that supports methods for parallel
training, especially for exact (i.e., interpolating) Gaussian processes.\cite{NoackExact2023}

For this study we use two simulation codes:  WarpX and Wake-T. WarpX
is a PIC simulation code that is used to model the problem in 3D
(the domain here is decomposed in the $z$-dimension along the beam). It uses MPI parallelism, where each MPI
task uses \revision{a CPU core for control logic and serial operations, and a GPU for intensive computational work.} The problem is configured to run using 48 GPUs, and it runs for
approximately 15--20 minutes on Perlmutter. The objective function of interest for WarpX is the
product of the beam divergence in the horizontal (\texttt{x}) and vertical (\texttt{y}) dimensions orthogonal to the beam; the Wake-T 
simulation is axisymmetric, and the beam divergence is taken in the \texttt{x} direction only. 

Wake-T (the Wakefield particle Tracker)\cite{FerranPousa2019} is used as a simplified serial code
that uses a Runge--Kutta solver to track the evolution of the beam electrons. Wake-T is used here to help configure \revision{the training algorithm used in the gpCAM generator} and to measure the overheads of libEnsemble using many workers.

\subsection{Motivation for Surrogate Model}

Building a surrogate for a plasma-based accelerator stage can help facilitate
numerous research and optimization tasks for future designs of high-energy
physics colliders that are based on the succession of hundreds to thousands of
such stages. Various simulation codes that are used for the modeling of stages,
from fast (e.g., the reduced dimensionality with the code Wake-T) to more detailed
and more computationally demanding (e.g., fully three-dimensional modeling with WarpX), can be
used to build such surrogates. A high-quality global surrogate model of Wake-T
and/or WarpX output (over some portion of parameter space) that can
sufficiently approximate a specific simulation scenario will enable
researchers to conduct extensive collider design experimentation and analysis
at a fraction of the time and computational cost. This is particularly
advantageous for collider designs where a wide array of scenarios and
objectives is to be considered. Such studies enable insights that might not
have been initially considered, including designs under various regimes of uncertainty.  Such surrogate models 
will likely be useful for innovation and robustness in particle accelerator applications.

\subsection{Online Gaussian Process Training with gpCAM}

The aim is to sample the parameter space \revision{in an online fashion to construct a Gaussian process surrogate model. A Gaussian process is a distribution over functions whose posterior (i.e., post-data) distribution is updated as new data (in the form of input simulation parameters and their associated evaluation) are obtained. One strategy for the online sampling uses the current posterior uncertainty and samples input parameters in areas of the parameter space where this uncertainty is largest. Gaussian processes depend on internal hyperparameters, which are not studied here, such as a length-scale hyperparameter whose value tends to determine how ``bumpy'' the resulting surrogate distribution is.}
Relative to offline/fixed-sample training, this kind of online learning, wherein the input parameters are sequentially selected based on a growing set of training data, can significantly reduce the number of simulation runs required to achieve a level of predictive accuracy \revision{of the surrogate}. 

The gpCAM generator is
initially run with random points drawn uniformly from the parameter space. The simulation output from this initial set is used to train the Gaussian process surrogate model. Following
this initial phase, the 5D parameter space is divided into a mesh of candidate points. The mesh used in the following case was
10 points in each dimension (hence $10^5$ points). The input \texttt{x} and the objective function \texttt{f} are used
to sample the posterior covariance from gpcAM at each of these candidate points. Noise is initialized to approximately 1\% of the mean \texttt{f}
value. 

With points ranked by their uncertainty (i.e., highest covariance), we select the top-ranked candidate point for sampling. To ensure sampling is spaced out across the parameter space, we proceed by identifying subsequent points that are at \revision{least a} distance \texttt{r}\revision{$>0$} away from previously selected points. This iterative selection is refined by initially setting a substantial value for \texttt{r} and gradually reducing it, allowing us to assemble a batch of simulation points that maintain a minimum separation of \texttt{r}. This strategy ensures both targeted exploration of high-uncertainty regions and spatial diversity within the batch. Once a desired batch size has been reached, simulation outputs are obtained for the entire batch, and the Gaussian process surrogate model is retrained by using the now larger set of training data. This iterative process then repeats, with inputs and their corresponding outputs being obtained in an online manner. libEnsemble has been used in a similar fashion for online Bayesian calibration.\cite{Surer2023}

\begin{figure*}
    \centering
    \includegraphics[width=0.49\linewidth]{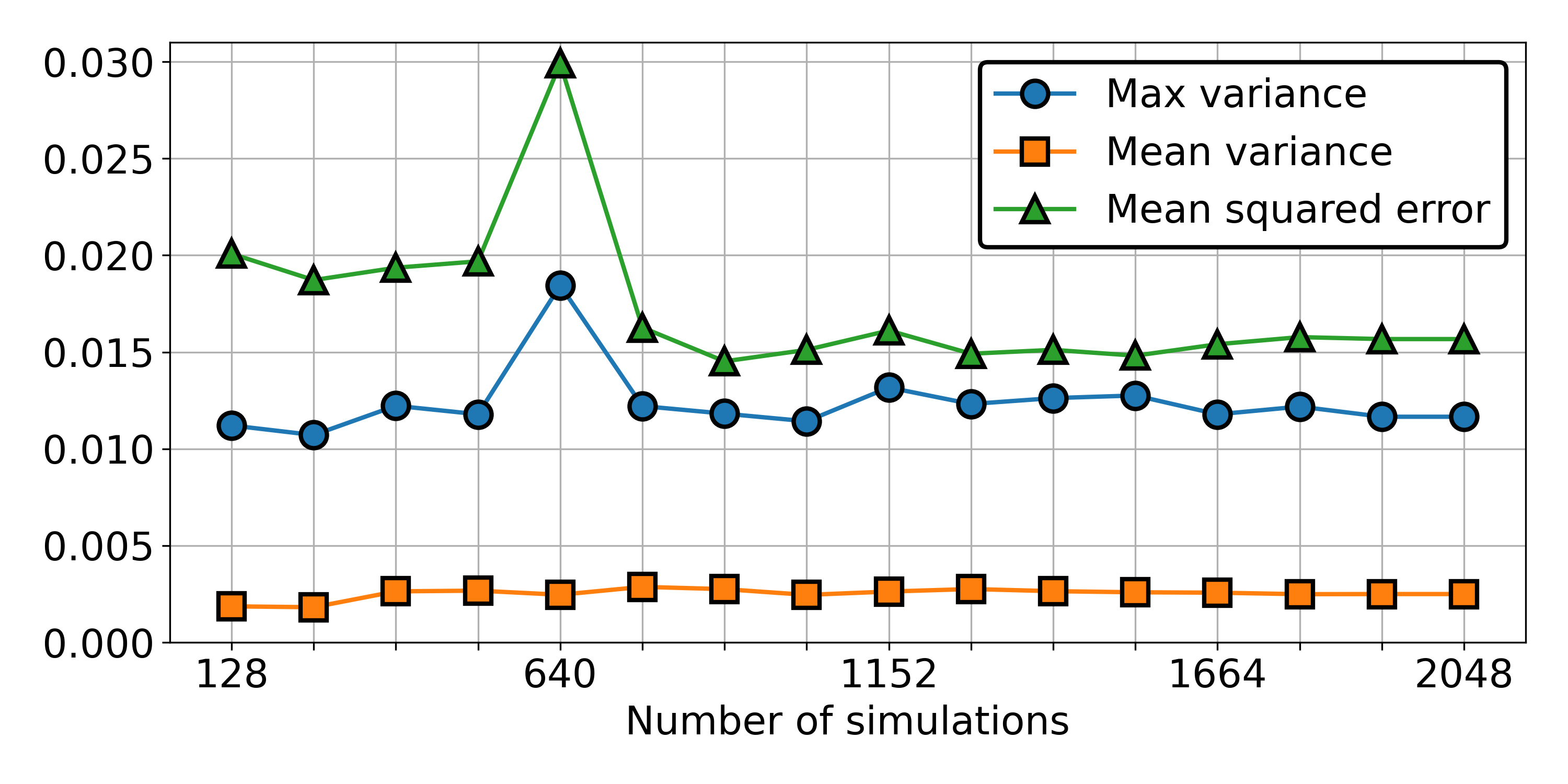}
    \includegraphics[width=0.49\linewidth]{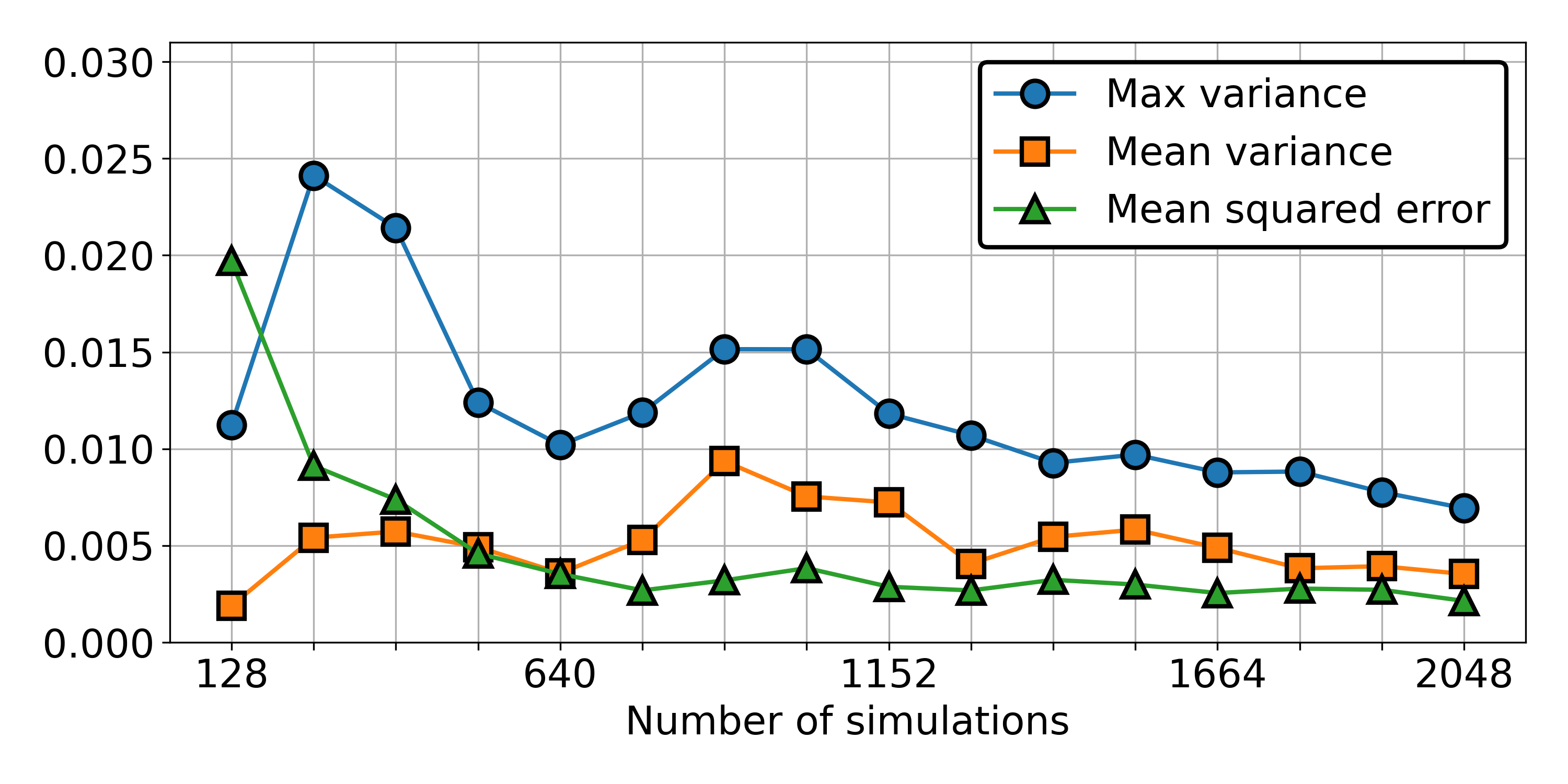}
    \caption{Behavior of Wake-T mean squared error as more training data is obtained (simulation input parameters chosen from a uniform random sample (left) and gpCAM (right) of parameter space).}
    \label{fig:random_pts_waket}
\end{figure*}

To assess whether the surrogate is improving, we monitor the \revision{Gaussian process variance (both the mean and maximum) on the set of} candidate points and the mean squared error (MSE) \revision{on} a test set (i.e., a separate set of sampled points). We do not necessarily expect to see monotonic improvement 
\revision{in any of the three metrics since the model may change significantly at a test or candidate point} 
if unexpected behavior of the objective is encountered
\revision{nearby.} 
Nonetheless, our expectation is that, with ongoing sampling and refinement of the model, both the variance indicators and the MSE will demonstrate a general downward trend, indicating enhanced model accuracy and reliability as more model training data is obtained.

\subsection{Training Details}

The platforms used were Perlmutter (NERSC) and Frontier (OLCF). Both platforms have demonstrated the scalability of WarpX and have been used extensively with libEnsemble.\cite{10046112} Perlmutter also has a CPU partition that is suitable for use with the Wake-T version.

To prepare the workflow and configure gpCAM, we first use a Wake-T axisymmetric model in place of the 3D WarpX. Since this runs on a single CPU core, it can be run either via libEnsemble's executor or directly in Python via a function call. For our purposes, a direct function call is sufficient and minimizes overhead. libEnsemble is run with the \texttt{mpi4py} communications option to spread workers across nodes. The use of 32 workers per node was found to scale well up to 2,048 workers (64 nodes). This workflow is used to configure the gpCAM generator and to measure and optimize model training, which can later be adapted to the WarpX workflow.

The full training of the Gaussian process takes nontrivial time, and this time increases as the number of evaluated points given to the model increases. Since we are working with batches, this means that worker resources will be delayed waiting for the Gaussian process to be retrained. A few approaches  can be used to improve this situation.

1. Only perform the necessary training at each step.
  One can assess the quality of the model and perform a full global training (120 iterations), a reduced global training (20 iterations), or a local training, depending on the model quality.

2. Employ different training methods.
  Markov chain Monte Carlo samples from a probability distribution based on constructing a Markov chain  may lead to more robust and reliable modeling in some scenarios.

3. Overlap training with evaluations. 
  libEnsemble supports asynchronous return of evaluations to the generator. This enables overlap of training based on returned evaluations, while concurrently continuing with further evaluations.

4. Use more resources for training.
  The gp2Scale feature of gpCAM supports parallel training via Dask.\cite{rocklin2015dask} This would need to be incorporated into the libEnsemble framework or use existing libEnsemble capabilities (this may entail using the libEnsemble executor within the generator to assign more resources for training). 
  
The first strategy enhances the efficiency of model training by making more judicious decisions on training the model. By default, gpCAM conducts a global training consisting of 120 iterations. We refer to this as full training. Once a robust model is established, reducing the number of iterations for global training becomes feasible. Additionally, transitioning to even faster local training (considering only the local Gaussian process hyperparameter region) becomes an option, offering a more targeted approach to refine the model.

The quality of the model can be assessed by comparing evaluation results predicted by the model with the actual results returned in the
last batch. This is similar to the use of test points to measure model quality. In this case we use the gpCAM function \texttt{rmse} (root mean
squared error), and we assess the quality of the model to assign training by comparing \texttt{rmse} against the standard deviation of
\texttt{f} via the logic shown below.

\begin{minted}[
frame=lines,
fontsize=\scriptsize,
]
{python}
rmse = gp.rmse(x_new, y_new)
gp.tell(all_x, all_y)
if rmse > 10.*np.std(all_y):
    gp.train(method='global', max_iter=120)
elif rmse > 2.*np.std(all_y):
    gp.train(method='global', max_iter=20)
else:
    gp.train(method='local')
\end{minted}

The above training thresholds, for this use case, provide sufficient training to ensure the model quality remains good enough to keep the variance (and error at test data) trending downward. The time spent training is reduced but would still become a bottleneck as larger numbers of evaluations are returned.

\subsection{Results for gpCAM with Wake-T}

\Cref{fig:random_pts_waket} shows the mean squared error at the test points at each iteration of the generator. The graphs also show the mean and maximum variance (posterior covariance) over the grid, as determined by gpCAM. These runs used batches of 128 concurrent evaluations (using 128 workers for running simulations and one worker for the persistent generator function).

\Cref{fig:random_pts_waket}(left) shows a run where uniformly random points were selected for evaluation and gpCAM was used only to evaluate the posterior mean (for mean squared error) at the test points and posterior variance over the grid of candidate points. This illustrates that while the mean variance of the model stays relatively stable, the maximum variance shows that areas of high uncertainty remain, indicating the existence of potentially underexplored regions. The mean squared error at the test points, with the exception of one spike, also remains roughly constant over this range of number of simulations performed. 

\Cref{fig:random_pts_waket}(right) shows results when the candidate points are selected using information from gpCAM via the online method described above. The results show that both the mean squared error at the test points and the maximum/mean variance over the grid of candidate points all trend downward over time. The results confirm the expectation that gpCAM is enabling efficient training of the surrogate model.

\subsection{Performance Considerations and Scaling}

Since the time taken to train the Gaussian process model and query the posterior covariance across the grid of points increases with the number of evaluations, ongoing work will explore the capabilities in gpCAM for more efficient training strategies and for parallelization of training using gp2scale.

During this study we consider using CRPS (Continuous Rank Probability Score) in place of the RMSE as a metric for evaluating model performance and, therefore, training requirements. While RMSE looks only at prediction error, which may be a limited measure in sparse scenarios, CRPS takes the model's uncertainty estimates into account and therefore may give a better indication of model confidence. Applying the same comparisons using CRPS in place of RMSE resulted in considerably more local retraining, which is significantly faster. However, there were also more frequent spikes in the variance suggesting this training may have been insufficient. We expect that refining the thresholds for different training options (relative to the standard deviation of the objective) will likely make CRPS a more efficient training strategy. This is left for future work. A comparison of Gaussian process hyperparameters used in gpCAM may also be used to make more informed training decisions.

Wake-T run times recorded inside the simulator, from 3,072 runs, average 137.1 seconds with a maximum of 144.7 seconds on the Perlmutter CPU partition using 32 workers per node. To assess the scaling of libEnsemble's infrastructure, timing is placed around the send/receive call inside the persistent generator function. This timing includes all overhead (including communication between the generator and manager and between manager and simulation workers in addition to time spent in simulations).

\begin{figure}
    \centering
    \includegraphics[width=4in]{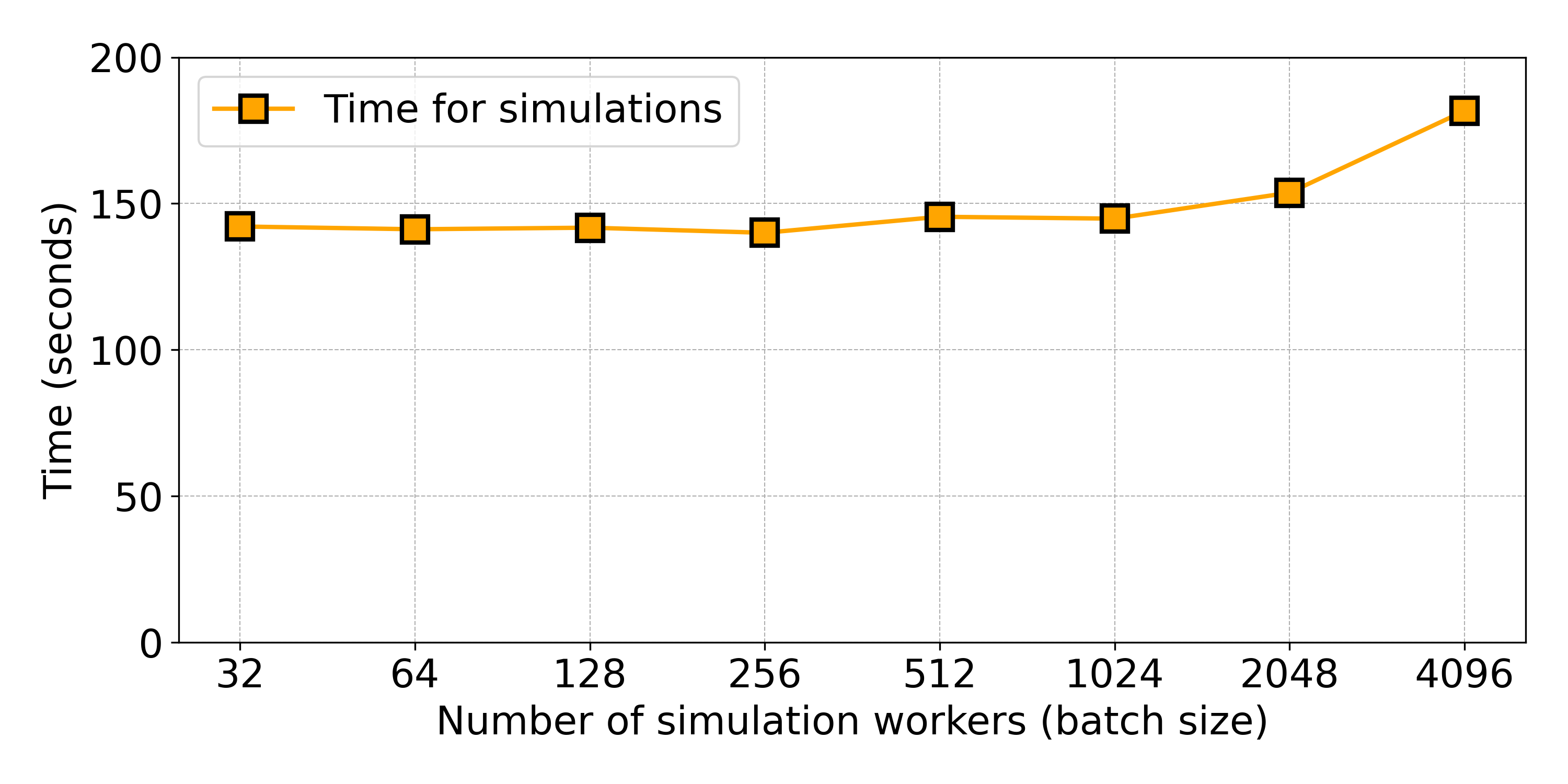}
    \caption{Wake-T weak scaling: Time for one batch of simulations measured by the generator. Note that the number of simulations
    is equal to the number of workers, hence showing weak
    scaling.}
    \label{fig:scaling_waket}
\end{figure}

\Cref{fig:scaling_waket} shows the timing obtained from the generator from one node (32 simulation workers) doubling up to 4,096 simulation workers. The data shows negligible overhead and nearly perfect speedup to 1,024 workers (98.2\% parallel efficiency relative to 32 workers), small overhead at 2,048 workers (92.5\% parallel efficiency), and significant overhead at 4,096 workers (78.2\%). This demonstrates the high-quality scalability of the \textit{mpi4py} communications method.

Note that the ability to run the generator on the manager has recently been added to libEnsemble and would likely reduce the communication overhead further.

\subsection{WarpX Study}

The WarpX version of the downramp simulation has been configured to run on 48 MPI tasks, each using one GPU. This setup is considered the minimum configuration that yields results with an acceptable physical accuracy.

\begin{figure}[b]
    \centering
    \includegraphics[width=0.5\linewidth]{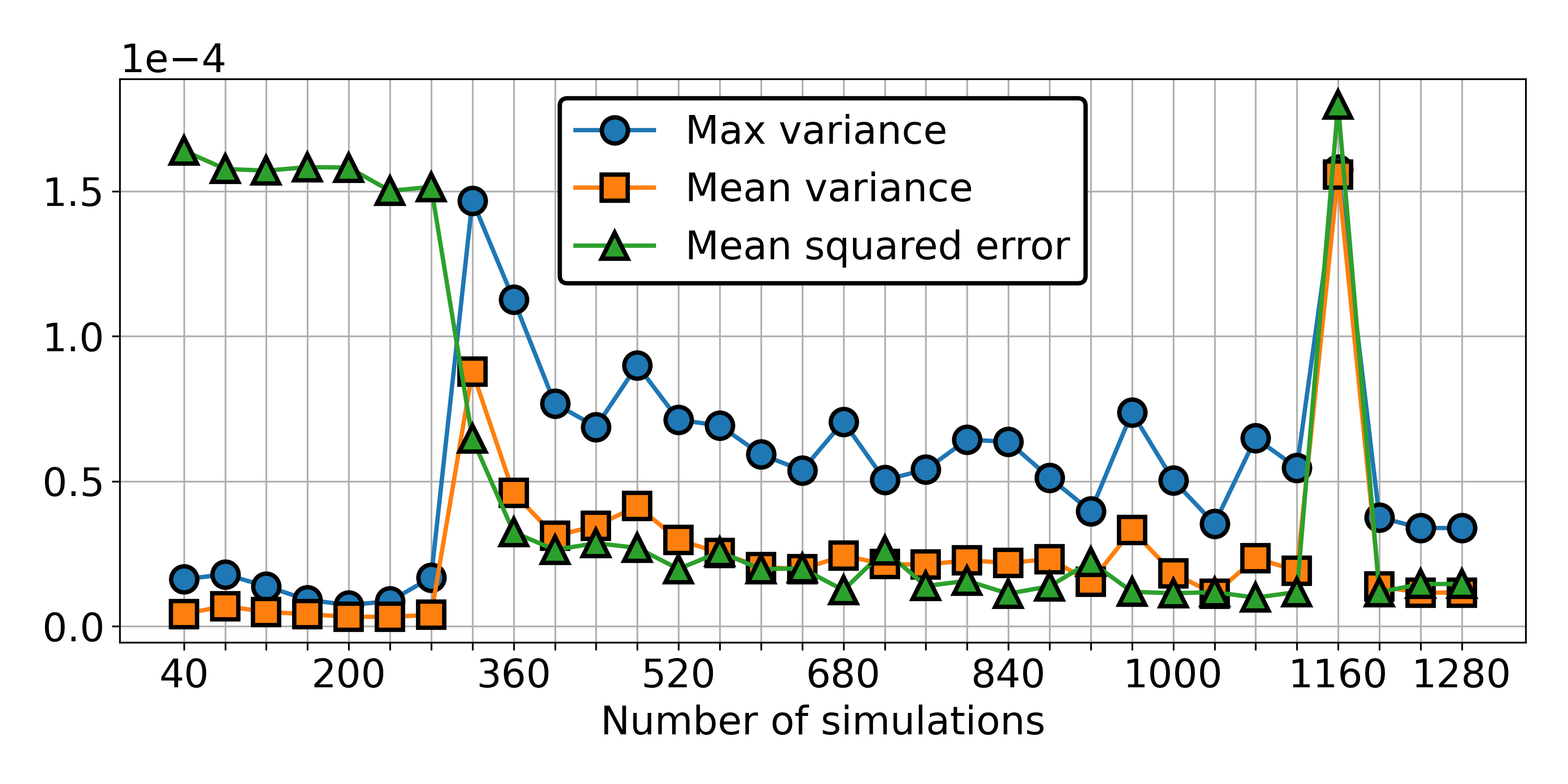}
    \caption{Behavior of WarpX mean squared error as more training data is obtained (simulation input parameters chosen by gpCAM process). }
    \label{fig:gpcam_pts_warpx}
\end{figure}

This means that on Perlmutter, each simulation, and hence each worker, uses 12 nodes (in contrast to 32 workers per node for Wake-T). If the number of parallel simulations was scaled to the same extent as the Wake-T study, 27 Perlmutters would be required. In fact, a full Perlmutter run for WarpX would use only 149 workers. On Frontier, there are eight GPUs per node, and hence each simulation requires only six nodes. On both systems, the WarpX run time is similar---approximately  800 seconds. On Perlmutter, however, about one run in 40 ran slowly and took up to 2,000 seconds. The reason for this is not clear, but the decision was taken to kill simulations that took longer than 1,000 seconds in order not to waste resources in a batched scenario. This is achieved through libEnsemble's executor interface, which has a portable kill function. The killed simulations return \texttt{NaNs}, which are excluded in the generator. Frontier, however, did not experience these slow runs. For these reasons, Frontier was favored for a large ensemble using WarpX. \Cref{fig:gpcam_pts_warpx} shows the results of running 32 iterations with a batch size of 40 on Frontier (a total of 240 nodes and 1920 GPUs).

\subsection{WarpX Results}

\begin{figure}[t]
    \centering
    \includegraphics[width=0.5\linewidth]{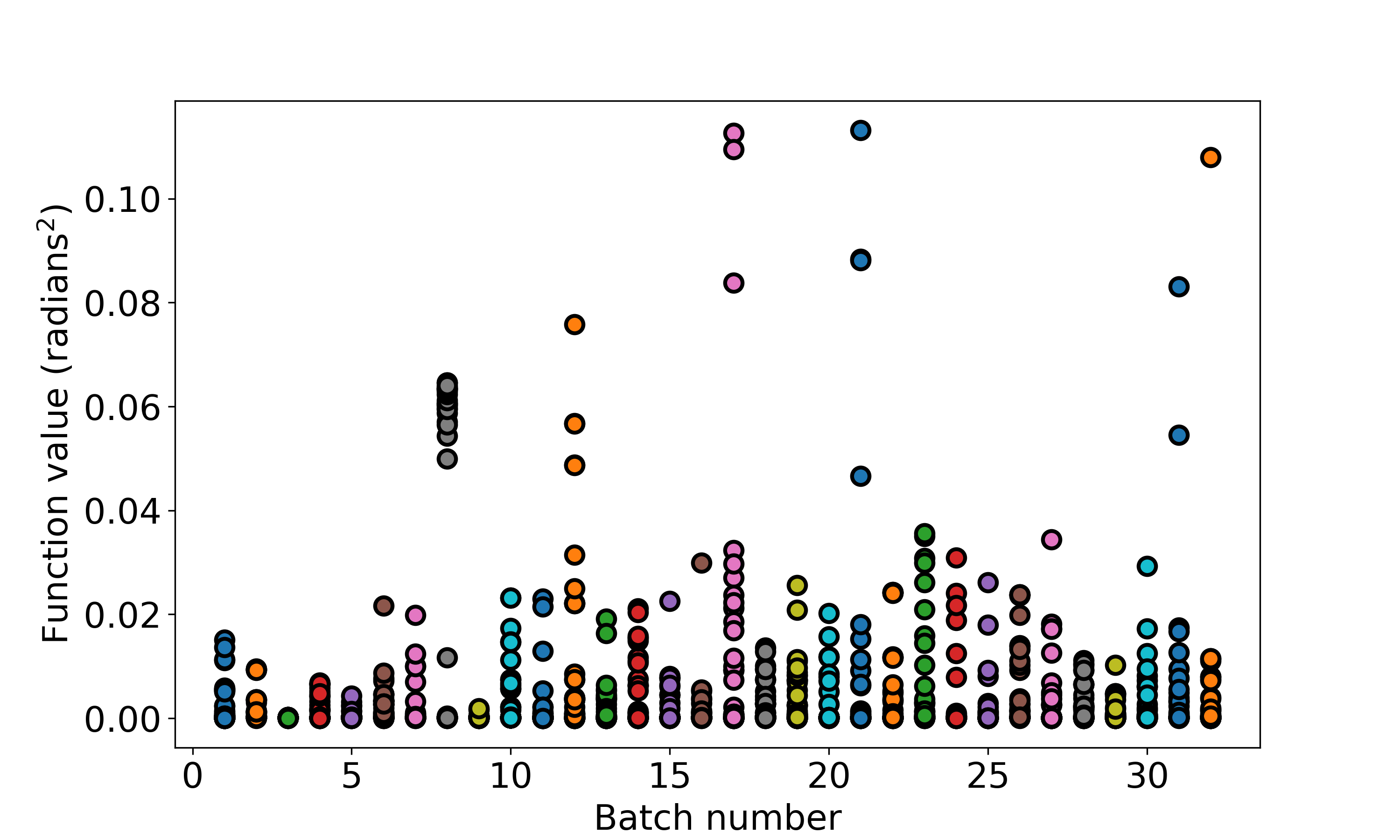}
    \caption{WarpX objective function value distribution by batch.}
    \label{fig:warpx_scatter}
\end{figure}

The mean squared error at the test points shows a downward trend with a steep improvement between batches seven and nine.
The model appears to start with fairly low uncertainty estimates; on processing, however, the eighth batch shows an increase in uncertainty but a reduction in error, suggesting an improvement of the surrogate. This corresponds with hitting some much larger objective values (shown in \Cref{fig:warpx_scatter}).

A spike occurs in the mean squared error and variance after 29 batches. This spike does not correspond to encountering obviously wayward values for the objective or to the RMSE for the last batch of points. However, examining the Gaussian process hyperparameters used in gpCAM shows that the length scale hyperparameter for the bump width is less than one-tenth of the range in this dimension, which may make the model oversensitive. 

The RMSE-directed training resulted in 10 full global training iterations, 20 smaller global training iterations, and one local training.

\begin{figure}[b]
    \centering
    \includegraphics[width=0.5\linewidth]{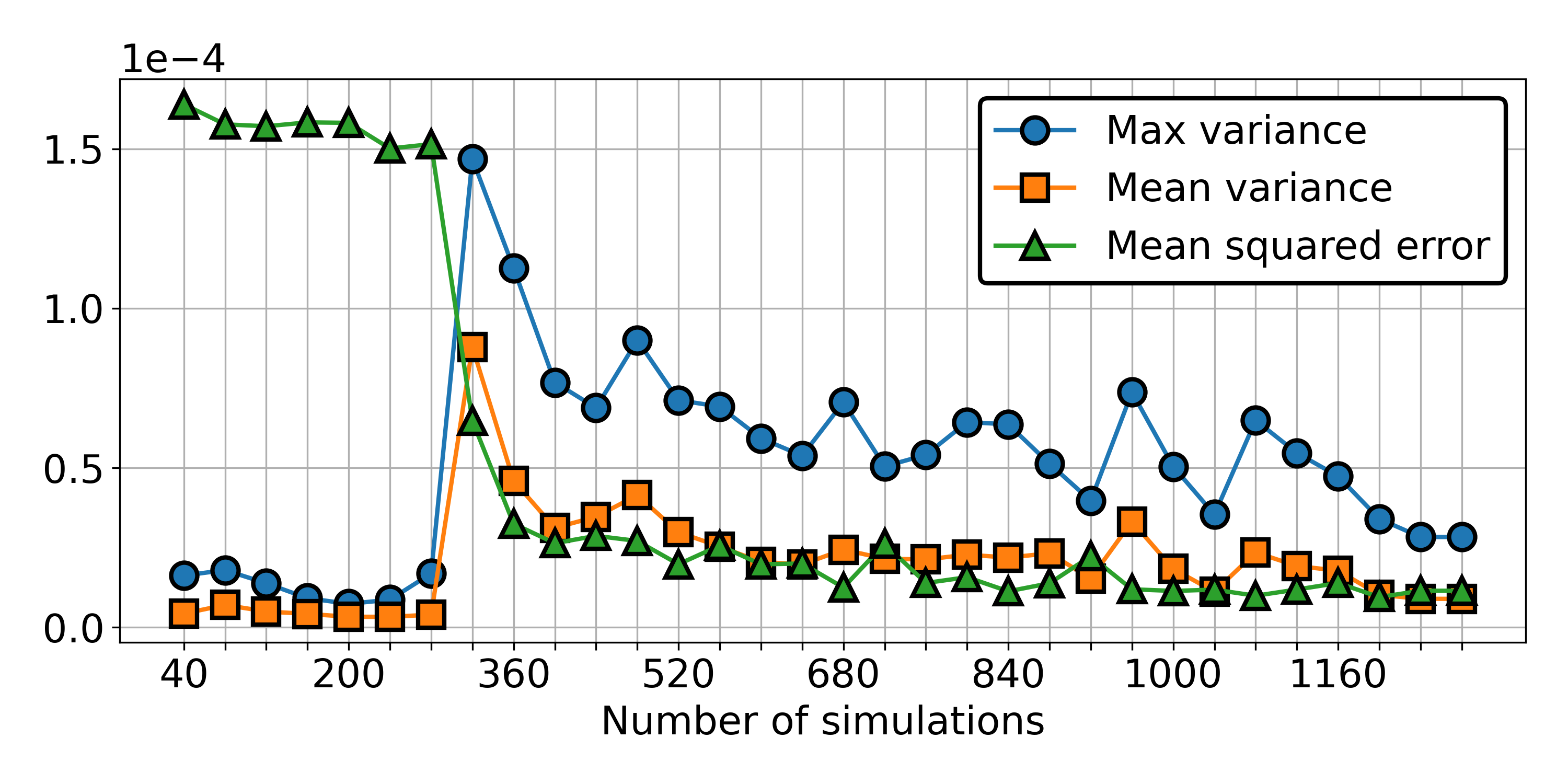}
    \caption{Behavior of WarpX mean squared error when gpCAM always performs global training  (simulation input parameters chosen by gpCAM process). }
    \label{fig:gpcam_pts_warpx_run2}
\end{figure}

The occurrence of the spike in variance (and mean squared error) was observed to be in close proximity to the single instance of local training.
Thus a second run was conducted on Frontier where the local training option was removed, thereby ensuring that either full or reduced global training was carried out by gpCAM. This second run, shown in \Cref{fig:gpcam_pts_warpx_run2}, does not result in the aforementioned spike. Future efforts will be directed toward developing strategies to more effectively handle anomalies in the Gaussian process hyperparameter values.

\begin{figure}[b]
    \centering
    \includegraphics[trim={40pt, 20pt, 40pt, 20pt}, clip, width=0.5\linewidth]{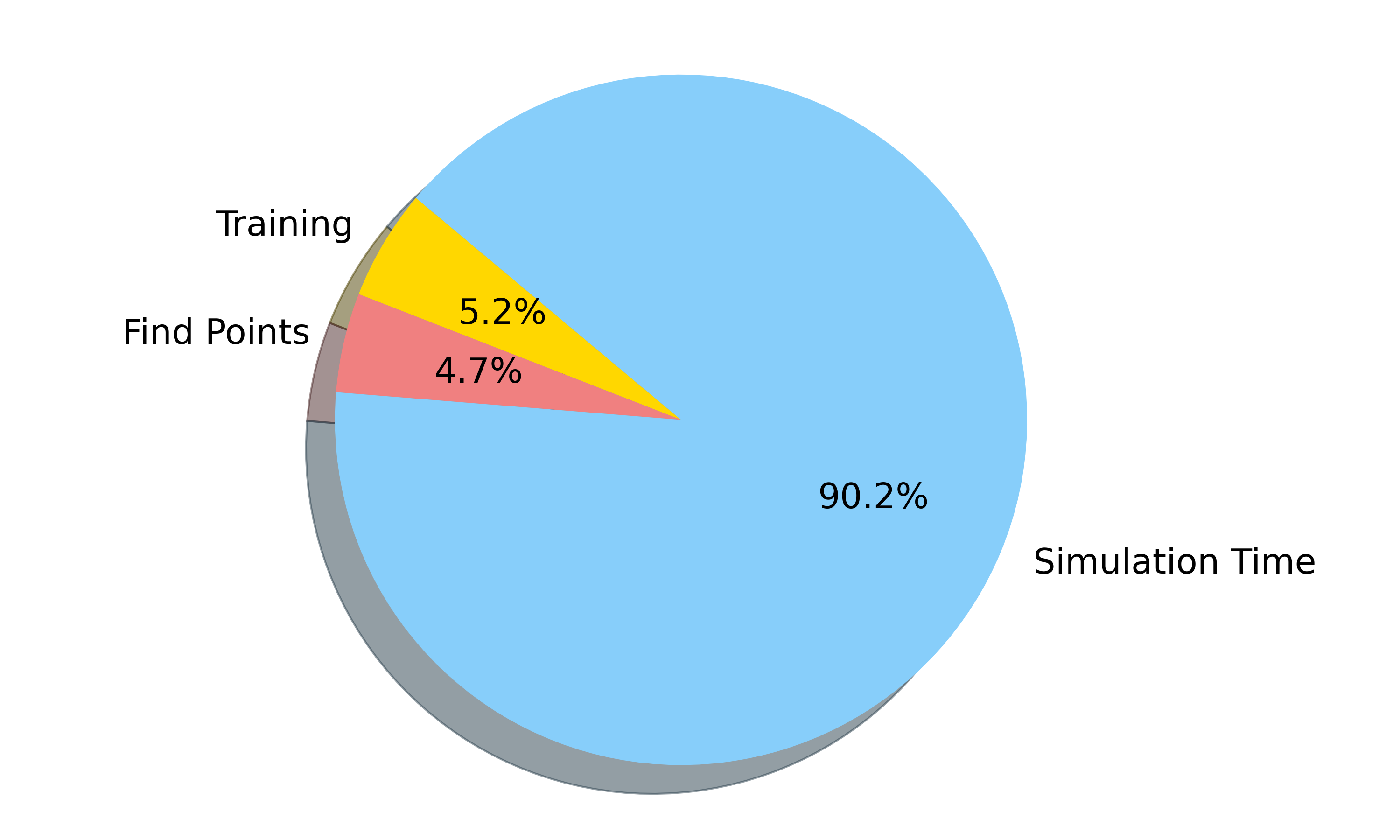}
    \caption{Time spent in training the model, deciding new points, and running WarpX simulations on Frontier ensemble}
    \label{fig:timing_breakdown_warpx}
\end{figure}

The overall time taken in this second run as measured from the generator function in training the model, selecting points (including measuring posterior covariance, sorting, and determining best points), and running simulations (including communication) is shown in \Cref{fig:timing_breakdown_warpx}. The overhead from manager-worker communications was negligible.

\subsection{Summary of Surrogate Case Study}

In summary, we were able to show progression toward a predictive surrogate model with both Wake-T and WarpX for the given use case. The lower-fidelity code (Wake-T) might not capture all of the physics, but it gives an initial understanding of the system response over the parameter space. WarpX captures more of the physics and while a sensitivity to training choices was observed, the mean squared error at the validation points trends downward, and several avenues for further configuration of the training protocol are available.

We demonstrated near-perfect weak scaling of libEnsemble's \texttt{mpi4py} communications to over a thousand workers on Perlmutter (a relative parallel efficiency of 98.2\% for 1,024 workers to 32 workers).

As long as the processing time spent in the generator remains minimal or is overlapped with simulations, extensive computational resources can be utilized for building such surrogate models. A full system run of the WarpX use case on Frontier would use 1,568 workers (9,408 nodes with six nodes per worker), which is within the scaling range demonstrated by the Wake-T use case, with a global parallel efficiency close to that of each individual simulation. 

libEnsemble has also been tested and shown to scale well on the Aurora supercomputer at the Argonne Leadership Computing Facility. Once WarpX has been successfully ported to Aurora, this is also expected to be a good system for further studies.

\section{Future Work}

libEnsemble is in active development and has a growing community of users. Its generalized interface
\revision{makes it well suited for} cutting-edge research across a large variety of disciplines and
experimental use cases. We consider the following to be especially worthy of further investigation.

\subsection{Using libEnsemble for Simulation Dispatch}

libEnsemble's in-the-loop generator facilitates close coupling of generators and simulators, especially in asynchronous scenarios. However, some users may wish to exploit libEnsemble's capabilities with an external generator. While doing so is already possible, a \textit{concurrent.futures} interface that allows libEnsemble to be used as a dispatch engine for simulations may better suit the needs of some users.
\revision{This interface would
also allow for easy integration with other tools. For example, as a Parsl\cite{parsl} executor.}

\subsection{Worker-Side Resource Management}

Another characteristic of the current resource implementation is that the manager controls resources and assigns these to workers. An alternative scheme is to manage resources on the worker side. Under this model, workers could use an alternative executor to dispatch tasks to a dedicated process that would assign resources and submit application runs. This method could leverage  much of the existing resource management and submission infrastructure. \revision{Additionally, it would be better suited for integration with external resource management tools such as Flux, as the detection and assignment of resources would happen in the same place. Finally, the method could work better with multisite workflows, where the manager is not directly able to detect the remote resources available to workers.}

\subsection{Improved Domain Support}

While libEnsemble's generalized interface is an advantage in adapting it to a wide variety of use cases, some users (e.g., the developers of Optimas) have  developed higher-level libraries that wrap libEnsemble for specific disciplines. As the amount of
artificial intelligence and machine-learning applications increases, libEnsemble can likewise serve that domain by exposing high-level \texttt{Model.train}
\texttt{test}, \texttt{ask}, \texttt{tell}, or other general functions.

\revision{Additionally, AI/ML may be further}  supported by improving multisite capabilities; the libEnsemble team is highly interested in coordinating ensembles
simultaneously across supercomputers and AI testbeds with special-purpose accelerators.

\subsection{Data Streaming}

libEnsemble's traditional data communications have been restricted to points, often single NumPy arrays that must be communicated
between the workers via the manager; in practice for large amounts of data, this approach is relatively slow and inefficient. This could be improved by sending a proxy or stream handle via the manager (e.g., using ProxyStore\cite{pauloski2023proxystore}), to transparently stream data directly between workers.

\subsection{Portable Generator Interface}

libEnsemble's user functions are currently constructed to be launched and processed by libEnsemble only.
This \revision{has} advantages such as ensuring that useful libEnsemble data structures are available within such
functions, guaranteeing they are portable across every libEnsemble version, and not forcing users to adhere to any
specific design patterns except input/output types. However, as the ecosystem of generator-algorithm studies and
upstream software that depends on libEnsemble grows, so does the request for libEnsemble to also launch and process third-party
generators. This is currently supported via writing wrapper user functions, although this wrapper code often fits design patterns that libEnsemble could handle on the user's behalf. 

Initial efforts indicate that many third-party generative interfaces consist of \texttt{.ask()} and \texttt{.tell()}
(or \texttt{.read()} and \texttt{.update()}) functions, as part of a class based generator. libEnsemble's workers could potentially interleave calling such functions with
sending/receiving corresponding data from libEnsemble's manager. Furthermore, if libEnsemble's traditional generators were also
reconstructed with that two-function paradigm, they would be available for use in other workflow packages.

\section{Acknowledgements}
This article was supported in part by the
PETSc/TAO activity within the U.S.\ Department of Energy's (DOE's) Exascale Computing
Project (17-SC-20-SC) and by the CAMPA, ComPASS, and NUCLEI SciDAC projects within
DOE’s Office of Science, Advanced Scientific Computing Research under contract numbers
DE-AC02-06CH11357 and DE-AC02-05CH11231.
We acknowledge contributions from David Bindel early in libEnsemble's
development. We thank Marco Garten, Remi
Lehe, Jean-Luc Vay, \'Angel Ferran Pousa,    and Axel Huebl for supplying Wake-T and WarpX use cases. We thank
Marcus Noack for assistance with gpCAM. 

This research used resources of the Oak Ridge Leadership Computing Facility, a DOE Office of Science User Facility supported under Contract
DE-AC05-00OR22725, and the National Energy Research Scientific Computing Center (NERSC), a DOE Office of Science User Facility located at Lawrence Berkeley National Laboratory and operated under Contract DE-AC02-05CH11231 using NERSC awards ASCR-ERCAP0028863 and HEP-ERCAP0027030.

\bibliographystyle{SageV}
\bibliography{libE.bib}

\section*{Biographies of the authors}
\textbf{Stephen Hudson}
  received his M.Sc.~(2001) in computer science from Oxford Brookes University.
  He is a software engineer and numerical library developer in LANS, the Laboratory for
  Applied Mathematics, Numerical Software, and Statistics, at Argonne National
  Laboratory. He specializes in high-performance computing.
  
\noindent \textbf{Jeffrey Larson}
  received his B.A.~(2005) in mathematics from Carroll College and his
  M.S.~(2008) and Ph.D.~(2012) in applied mathematics from the University of
  Colorado Denver. He is a computational mathematician in LANS, the Laboratory for
  Applied Mathematics, Numerical Software, and Statistics, at Argonne National
  Laboratory. He studies algorithms for optimizing computationally expensive
  functions arising in quantum information science, fusion power plant and particle accelerator
  design/operation, and vehicle routing.
  
\noindent \textbf{John-Luke Navarro}
  received his B.A.~(2018) in computer science from Andrews University. He is a
  software engineer in LANS, the Laboratory for Applied Mathematics, Numerical
  Software, and Statistics, at Argonne National Laboratory.
  
\noindent \textbf{Stefan M.\ Wild} received B.S.\ and M.S.\ degrees in applied
  mathematics from the University of Colorado in 2003 and a Ph.D.\ degree in
  operations research from Cornell University in 2009. He is a senior
  scientist and director of the Applied Mathematics
  and Computational Research Division at Lawrence Berkeley National
  Laboratory. He is an adjunct faculty in Industrial Engineering and
  Management Sciences at Northwestern University. His research interests
  include the development of formulations, algorithms, and software for
  challenging mathematical optimization problems.

\vfil
\framebox{\parbox{.90\linewidth}{\scriptsize The submitted manuscript has been created by UChicago Argonne, LLC, Operator of Argonne National Laboratory (``Argonne''). Argonne, a U.S.\ Department of Energy Office of Science laboratory, is operated under Contract No.\ DE-AC02-06CH11357.  The U.S.\ Government retains for itself, and others acting on its behalf, a paid-up nonexclusive, irrevocable worldwide license in said article to reproduce, prepare derivative works, distribute copies to the public, and perform publicly and display publicly, by or on behalf of the Government.  The Department of Energy will provide public access to these results of federally sponsored research in accordance with the DOE Public Access Plan \url{http://energy.gov/downloads/doe-public-access-plan}.}}

\end{document}